\documentclass[aps,prb,reprint,superscriptaddress, amsmath,amssymb]{revtex4-1}

\usepackage{graphicx}
\usepackage{dcolumn}
\usepackage{bm}

\begin{document}


\title{Algorithms and image formation in orbital tomography}

%
%
%

\author{Pavel Kliuiev}
\email{kliuiev@physik.uzh.ch}
\affiliation{Department of Physics, University of Zurich, Zurich, Switzerland}
\author{Tatiana Latychevskaia}
\altaffiliation{Current address: Institute of Physics, \'Ecole Polytechnique F\'ed\'erale de Lausanne, Lausanne, Switzerland}
\affiliation{Department of Physics, University of Zurich, Zurich, Switzerland}
\author{Giovanni Zamborlini}
\author{Matteo Jugovac}
\affiliation{Peter Gr\"unberg Institute (PGI-6), Forschungszentrum J\"ulich GmbH, 52425 J\"ulich, Germany}
\author{Christian Metzger}
\author{Manuel Grimm}
\author{Achim Sch\"{o}ll}
\affiliation{University of W\"{u}rzburg, Experimental Physics VII, 97074 W\"{u}rzburg, Germany}
\author{J\"urg Osterwalder}
\author{Matthias Hengsberger}
\author{Luca Castiglioni}
\email{luca.castiglioni@physik.uzh.ch}
\affiliation{Department of Physics, University of Zurich, Zurich, Switzerland}

\date{\today}

\begin{abstract}
Orbital tomography has recently been established as a technique to reconstruct molecular orbitals directly from photoemission data using iterative phase retrieval algorithms. In this work, we present a detailed description of steps for processing of the photoemission data followed by an improved iterative phase retrieval procedure and the interpretation of reconstructed two-dimensional orbital distributions. We address the issue of background subtraction by suggesting a signal restoration routine based on the maximization of mutual information algorithm and solve the problem of finding the geometrical center in the reconstruction by using a tight-centered object support in a two-step phase retrieval procedure. The proposed image processing and improved phase retrieval procedures are used to reconstruct the highest occupied molecular orbital of pentacene on Ag(110), using photoemission data only. The results of the reconstruction agree well with the density functional theory simulation, modified to comply with the experimental conditions. By comparison with photoelectron holography, we show that the reconstructed two-dimensional orbital distribution can be interpreted as a superposition of the ``in-focus'' orbital distribution evaluated at the $z=0$ plane and ``out-of-focus'' distributions evaluated at other $z=\mathrm{const}$ planes. Three-dimensional molecular orbital distributions could thus be reconstructed directly from two-dimensional photoemission data, provided the axial resolution of the imaging system is high enough.
\end{abstract}

\pacs{Valid PACS appear here}
\maketitle


\section{\label{sec:Introduction}Introduction}

Orbital tomography provides means for the reconstruction of the amplitude and phase distribution of molecular orbitals solely from angle-resolved photoelectron spectroscopy (ARPES) data~\cite{puschnig:2009, puschnig:2013, lueftner:2014, weiss:2015, kliuiev:2016}. 
In the absence of final state scattering~\cite{dauth:2014}, e.g. when the photoemission signal is recorded from a well ordered monolayer or multilayer of organic molecules consisting of light atoms (H, C, N, O), a plane wave $\propto e^{i\textbf{k}_\mathrm{f}\textbf{r}}$ can be used to describe the photoemission final state~\cite{goldberg:1978, puschnig:2009, puschnig:2013}. This approximation makes it possible to relate the experimental photoelectron angular distribution (PAD) to the squared modulus of the Fourier transform $\mathcal{F}$ of the initial state wave function~\cite{puschnig:2009, puschnig:2013}:
\begin{equation}
I(\textit{\textbf{k}}_{\mathrm{f}\parallel}, E_{\mathrm{kin}})\propto|\textit{\textbf{A}}\cdot \textit{\textbf{k}} _{\mathrm{f}} |^2|\mathcal{F}\{\psi_i (\textit{\textbf{k}}_{i\parallel}, \textit{\textbf{r}})\}|^2.
\label{eq:photocurrent}
\end{equation}
In Eq.~\ref{eq:photocurrent}, $\textit{\textbf{k}}_{\mathrm{f}}$ is the photoelectron wave vector of the final state, $\textit{\textbf{A}}$ is the vector potential of the electromagnetic field and $I(\textit{\textbf{k}}_{\mathrm{f}\parallel}, E_{\mathrm{kin}})$ is the photocurrent recorded at the kinetic energy $E_\mathrm{kin}$. The photocurrent is obtained by summation over all electronic transitions from occupied initial states, $\psi_i$, to the final state, $\psi_\mathrm{f}$, characterized by the corresponding wave vector components $\textit{\textbf{k}}_{i\parallel}$ and $\textit{\textbf{k}}_{\mathrm{f}\parallel}$ parallel to the surface, respectively. We note that in practice, experimental data must often be deconvoluted~\cite{puschnig:2011} before Eq. \ref{eq:photocurrent} can be used. Also, the derivation of Eq. \ref{eq:photocurrent} requires making several approximations, which were in detail addressed by Dauth et al.~\cite{dauth:2014}. Namely, by assuming that the correlation between the photoemitted electron and the remaining ones is negligible and $\textit{\textbf{A}}$ is constant in space, the many particle matrix element becomes related to the Dyson orbital~\cite{mignolet:2013} and thus can be written in the form of a single-particle matrix element. In the plane wave approximation, the ARPES intensity becomes proportional to the Fourier transform of the Dyson orbital~\cite{dauth:2014}. Dauth et al.~\cite{dauth:2011} showed that orbitals obtained using self-interaction-free Kohn-Sham density functional theory are the best single-particle approximates to the Dyson orbitals probed by photoemission.

Provided the phase distribution of the photoelectron wave in the detector plane is known, the initial state wave function can be reconstructed by computing the inverse Fourier transform of the square root of the ARPES data. The phase distribution can be guessed from the parity of the wave function~\cite{puschnig:2009} or from dichroism measurements~\cite{wiessner:2014}. Without such knowledge, the phase distribution in the detector plane can be obtained iteratively by either confining the wave function to a rectangular box whose dimensions are determined by the van-der-Waals size of the molecule~\cite{lueftner:2014} or in a more robust manner~\cite{kliuiev:2016} by employing phase retrieval algorithms~\cite{fienup:1978, fienup:1982, marchesini:2003, harder:2010} used in coherent diffraction imaging~\cite{miao:1999}. These algorithms were optimized for reconstruction of complex-valued object distributions~\cite{harder:2010} and require only a rough estimate of the size of the object to ensure the fulfillment of the oversampling condition~\cite{miao:1998, miao:2003}. No \textit{prior} information about the shape of the object is needed, as the object support~\cite{fienup:1982} is found in the course of the reconstruction solely from experimental data using the shrinkwrap algorithm~\cite{marchesini:2003}. We note that the correct reconstruction in case of complex-valued object distributions is possible only when the support is (i) tight enough and (ii) belongs to one of a number of special types, which include supports consisting of separated parts and with no parallel sides~\cite{fienup:1987}. The use of the shrinkwrap algorithm~\cite{marchesini:2003} ensures the reduction of the number of parallel sides and the tightness of the support, as the support envelope shrinks progressively around the object distribution in the course of the reconstruction until the algorithm converges. The estimation of the object support together with the object distribution itself is particularly important if orbital tomography aims at reconstruction of orbital distributions of excited states whose shape and symmetry properties might be difficult to predict.

In orbital tomography, the molecules are adsorbed on a single crystal substrate, which ensures that the molecules are well ordered. It is also required that the interaction between molecular species and the substrate is weak~\cite{puschnig:2009, puschnig:2013, lueftner:2014}. This allows to reconstruct the orbitals of quasi ``free-standing'' molecules and then directly compare them with the gas phase simulations~\cite{puschnig:2009, puschnig:2013, lueftner:2014}. However, the PADs, recorded at binding energies of molecular states, might contain not only features due to photoemission from ``pure'' molecular states, but also some signal from the substrate that might be present at the same binding energies~\cite{ziroff:2010, lueftner:JESRP:2014, puschnig:2011, stadtmueller:2012}. 
Since such substrate features do not originate from the molecular states to be probed, such substrate signal should not be present in the PAD data for orbital reconstruction. 
This issue can be solved either (i) by choosing a substrate with a low density of states at the respective binding energies of molecular states~\cite{puschnig:2009, puschnig:2013,  lueftner:2014, weiss:2015, kliuiev:2016}, which however is not always possible or desirable, or (ii) by applying a fitting procedure~\cite{lueftner:JESRP:2014}. Unfortunately, these fitting algorithms may not eliminate all spurious features completely~\cite{lueftner:JESRP:2014}. In addition, they require acquisition of complete PADs throughout a broad range of binding energies at a sufficiently dense sampling rate. And since the phase retrieval algorithms perform better when the data were acquired with higher statistics, the acquisition time may therefore increase substantially. To solve this issue, we suggest using a signal restoration procedure consisting of registration of signal and background data via maximization of mutual information~\cite{maes:1997, mattes:2001}. We show that this procedure effectively removes the principal background features. As a result, instead of recording data at a broad range of binding energies with high statistics and applying the fitting procedure, we suggest performing a fast survey followed by acquisition of PADs with high statistics only at binding energies of the identified molecular states. We note that the suggested procedure can be applied only in case of weak substrate-molecule interaction and can extract only the most relevant molecular features lying above the noise level.

By applying this signal restoration procedure to the ARPES data recorded from the highest occupied molecular orbital (HOMO) of pentacene on Ag(110) and employing the phase retrieval algorithms~\cite{fienup:1978, fienup:1982, marchesini:2003, harder:2010, kliuiev:2016}, we reconstruct both the amplitude and phase distribution of the orbital without any symmetrization of the data or prior information about parity or shape of the wave function. The reconstruction is done using an improved version of our reconstruction procedure~\cite{kliuiev:2016}, by adding an additional refinement step, allowing for the precise determination of the geometrical centres of the reconstructed orbital distributions.

Finally, we compare the results of the reconstruction with the density functional theory (DFT) simulation of the pentacene molecule in free space. By using the imaging integral of Barton~\cite{barton:1988}, we interpret the reconstructed orbital distributions as a superposition of ``in-focus'' and ``out-of-focus'' contributions from the three dimensional (3D) orbital and discuss the possibility of the 3D reconstruction solely from a single set of two-dimensional (2D) experimental data.

We note, however, that orbital tomography is not the only method for visualization of molecular orbitals. The article by Schwarz~\cite{schwarz:2006} provides an overview of alternative imaging techniques, such as scanning tunneling microscopy or high harmonic spectroscopy and discusses subtle questions of quantum mechanics in view of the interpretation of reconstructed orbital distributions.

\section{\label{sec:Methods}Methods}

\subsection{\label{sec:Acquisition of ARPES data}Acquisition of ARPES data}

In principle, ARPES data suitable for orbital tomography can be acquired with any photoelectron spectrometer capable of systematically scanning a large range of electron emission angles. In practice, this can be achieved either by rotating the sample with a suitable manipulator~\cite{Greber:1997aa} or by using an angle-resolving electron analyzer with a large acceptance angle~\cite{Kromker:2008aa}. One particularly efficient instrument is the photoemission electron microscope (PEEM)~\cite{schneider:2012, tusche:2015} available e.g. at the NanoESCA beamline at the synchrotron radiation facility Elettra~\cite{Carsten-Wiemann:2011aa}. We use ARPES data recorded with this apparatus to illustrate the image processing and reconstruction algorithms. Fig. \ref{fig:figure1} (a) shows the PAD recorded from the HOMO of pentacene on Ag(110) at a binding energy of $E_\mathrm{b}=1.2$ eV with p-polarized light of $40$~eV photon energy. The crystal was prepared according to standard procedures~\cite{feyer:2014} and pentacene molecules were deposited from a home-built Knudsen cell. The PAD was obtained by averaging over 50 geometrically aligned raw data sets. The details of the experimental geometry and alignment of raw data are given in the Appendix. In Fig. \ref{fig:figure1}(a), the broad blobs are attributed to photoemission from the molecular state, while the narrow sharp features crossing them originate from the sp-bands of the substrate and have to be eliminated prior to application of the phase retrieval algorithm.

\subsection{\label{sec:Subtraction of substrate background}Subtraction of substrate background}

\begin{figure*}
\includegraphics{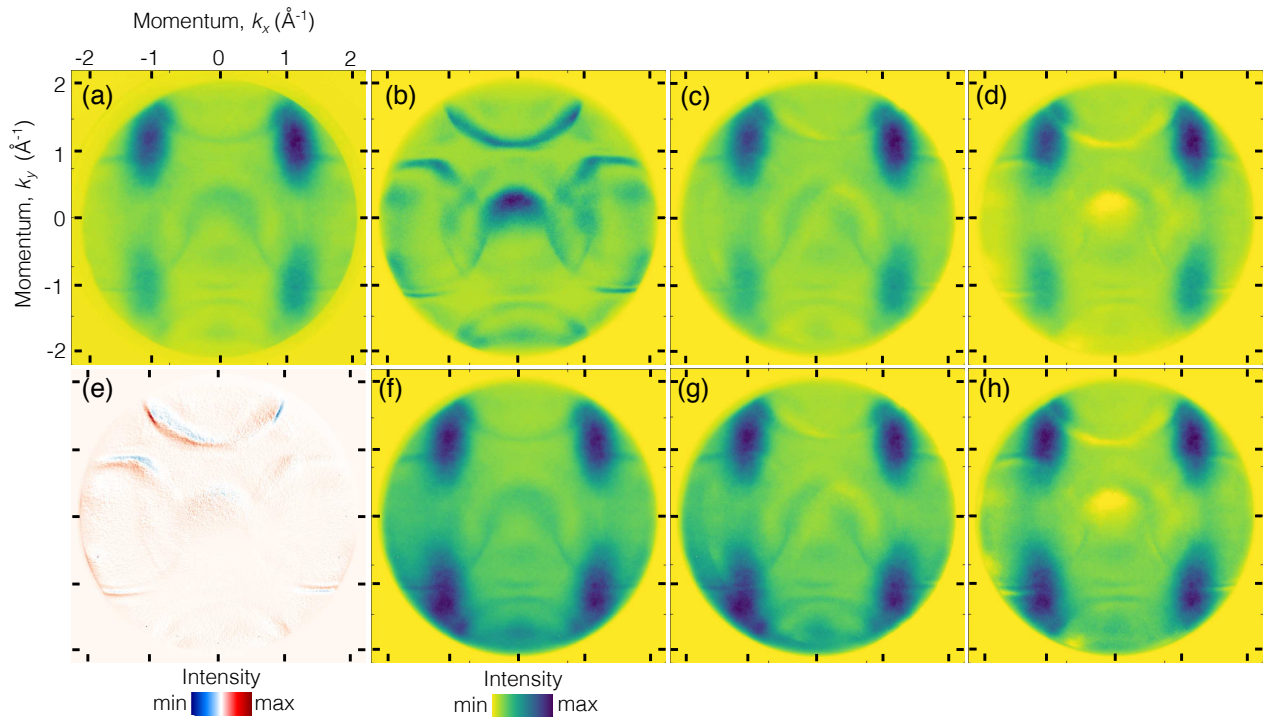}
\caption{\label{fig:figure1} (a) PAD recorded from a submonolayer of pentacene molecules at 1.2 eV binding energy. Sum of 50 geometrically aligned data sets. (b) Ag(110) substrate background at 1.2 eV binding energy. (c) The same PAD as in (a), but after the subtraction of the background in (b) registered with the PAD in (a) using the intensity interpolation method, in which only a mutual translation was accounted for. (d) The same PAD as in (a), but after the subtraction of the background in (b) registered with the PAD in (a) using the maximization of mutual information algorithm via iterative application of affine transformation to the PAD in (b). (e) Difference between the PAD in (b) and the same PAD, but registered with the PAD in (a) using the maximization of mutual information algorithm via iterative application of affine transformation to the PAD in (b). (f-h) The same PADs as in (a,c,d), but normalized by the $|\textit{\textbf{A}}\cdot\textit{\textbf{k}} _{\mathrm{f}}|^2$ factor.}
\end{figure*}

The removal of substrate bands from the signal data cannot be done by direct subtraction of the two images, because the background features in Fig. \ref{fig:figure1}(b) were sampled at slightly different $k_{\textrm{f},\parallel}$ values than in Fig. \ref{fig:figure1}(a). This deformation arose because neither the position of the sample with respect to the PEEM electron optics nor the PEEM settings were identical in both cases, which lead to changes in the field of view and thus to distortion of the substrate features in Fig. \ref{fig:figure1}(a) compared to those in Fig. \ref{fig:figure1}(b). Consequently, the background data shown in Fig. \ref{fig:figure1}(b) had to be brought into spatial registry with the corresponding features of the signal data shown in Fig. \ref{fig:figure1}(a) before subtracting the substrate contribution. 

Mutually shifted images can be aligned by means of the intensity interpolation method~\cite{tian_huhns:1986}, in which the normalized cross-correlation of the two images acts as a similarity metrics and has its maximal value at the position of the mutual displacement~\cite{brown:1992, zitova:2003}. Generalized versions of this method can also be used to align images distorted by affine transformations, perspective changes or optical aberrations~\cite{berthilsson:1998, hanaizumi:1993, simper:1996, zitova:2003}. 

First, we used the intensity interpolation method~\cite{tian_huhns:1986} to align the background data shown in Fig. \ref{fig:figure1}(b) with the signal data shown in Fig. \ref{fig:figure1}(a). The procedure was identical to the one used for alignment of 50 raw PADs and is described in the appendix. This procedure did not lead to a proper alignment, because upon subsequent subtraction of the registered background data from the signal image, many background features remained present, as shown in Fig. \ref{fig:figure1}(c). One reason for the failure of the intensity interpolation method is that it considers only translations and does not account for other possible types of distortions in order to properly map the background features in Fig. \ref{fig:figure1}(b) to those in Fig. \ref{fig:figure1}(a). Another reason for poor alignment with the intensity interpolation method can be attributed to the main drawback of the registration methods relying on the cross-correlation as the similarity metric, i.e. to their sensitivity to changes in the image intensity, introduced by noise or different imaging conditions~\cite{brown:1992, viola:1997, zitova:2003}. To account for these shortcomings, we assumed the transformation mapping the features in Fig. \ref{fig:figure1}(b) to those in Fig. \ref{fig:figure1}(a) to be affine, i.e. apart from translation, it accounted for rotation, scale and shear. The alignment was done via iterative application of affine transformations to the background image. Also, we employed mutual information (MI)\cite{collignon:1995, studholme:1995, wells:1996, maes:1997, viola:1997} as a more general similarity metric representing a measure of statistical dependence of two images acquired under varying imaging conditions. At each iteration, the pixel intensity values of the signal, $I$, and the background, $I_0$, each sampled at $N\times N=540\times 540$ pixels, were represented by two histograms divided into $N_\mathrm{bins}=127$ bins. The number of bins was determined by Scott's rule with the skewness factor~\cite{legg:2013} applied to the histogram of the background image. The quality of alignment was assessed by computing the MI metric as
\begin{equation}
\mathcal{S}(I,I_0)=\sum\limits_{i,j=1}^{N_\mathrm{bins}} p'(i,j)\log\frac{p'(i,j)}{p(i)p_0(j)},
\end{equation}
where $p(i)$, $p_0(j)$ and $p'(i,j)$ are the marginal signal, marginal background and joint probability distributions computed as continuous estimates using zero-order and cubic spline Parzen windows~\cite{thevenaz:1997}, as described in the Mattes' algorithm~\cite{mattes:2001}.

The optimization process, i.e. the iterative application of affine transformations, was driven by the one-plus-one evolutionary algorithm~\cite{styner:2000}. The optimization was done in $n=4$ cycles, each consisting of 1000 iterations. In the first cycle, the linear number of pixels was set to $N/(2n)$ in each dimension and then increased at the beginning of each new cycle by a factor of 2, until it was again equal to $N$ in the last cycle. This procedure allowed for a gradual refinement of the optimization results, until the algorithm converged after $200-250$ iterations in the last cycle. 

The mutual information of the background and signal data registered via maximization of mutual information and iterative application of affine transformations was equal to 0.3154, given that the mutual information of the two identical sets is unity. The registered background was subtracted from the signal image and the resulting background free image, $I_1$, is shown in Fig. \ref{fig:figure1}(d). The difference between the background data after and before registration is shown in Fig. \ref{fig:figure1}(e). It is seen already by a visual inspection of Fig. \ref{fig:figure1}(d) that the subtraction of the background features by means of the maximization of mutual information algorithm was more effective than that by means of the intensity interpolation, as most of the background features become suppressed upon subtraction. More quantitative insight was gained by computing the mutual information metric. The MI of the raw background and signal data was equal to 0.2152. In case of data registered by the intensity interpolation method, in which only a translation was taken into account, the MI was equal to 0.2157. In case of registration via maximization of mutual information, the metric was 0.3154. Thus, registered background data contained $\approx$ 46$\%$ more information about the background features in the signal data, compared to the image registration with the intensity interpolation method.

\begin{figure*}
\includegraphics{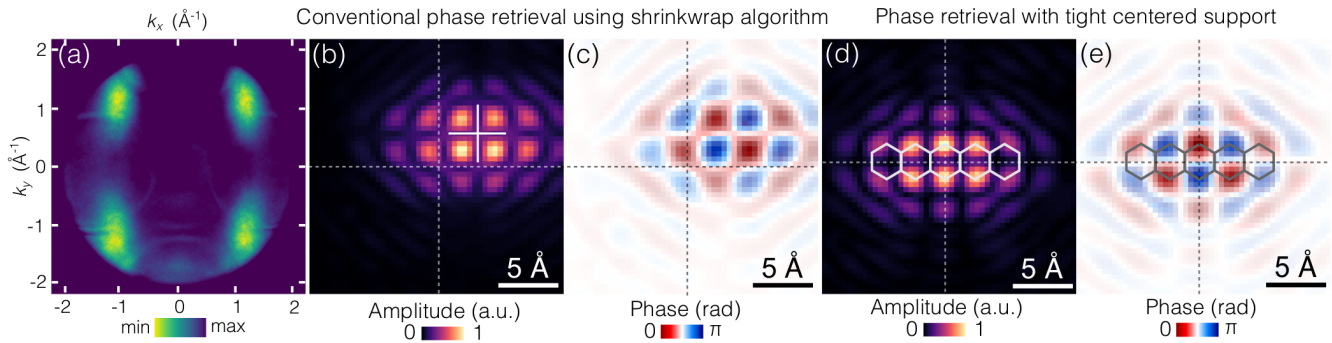}
\caption{\label{fig:figure2} Reconstruction of the HOMO of pentacene on Ag(110). (a) Final PAD used as an input for the phase retrieval algorithm, obtained after subtraction of the mean intensity value from each pixel in the PAD shown in Fig. 1(h). (b) Amplitude and (c) phase distributions reconstructed with the shrinkwrap algorithm using the uncentered support constraint. (d) Amplitude and (e) phase distributions reconstructed using the centered tight support obtained from the amplitude distribution in (b). The transparency of the phase images is weighted with the corresponding amplitude values for illustration purposes. The crossing dotted lines mark the geometrical centres of the computational domains. Images (b-e) are $70\times70$ pixels sections cut out from $2000\times2000$ pixels reconstructed images.}
\end{figure*}

To account for the modulation of photoemission data due to the angular dependency of the $|\textit{\textbf{A}}\cdot\textit{\textbf{k}} _{\mathrm{f}}|^2$ factor in Eq.~\ref{eq:photocurrent}, the PADs in Fig. \ref{fig:figure1}(a,c,d) were normalized by this factor. The details of the normalization procedure are given in the Appendix. The resulting distributions are shown in Fig. \ref{fig:figure1} (f-h). Obviously, while the PAD in Fig. \ref{fig:figure1}(h) lost most of the sharpest background features, their diffuse remnants were still present around the broad blobs. To eliminate the remaining quasi constant background, we subtracted the mean intensity value of the entire image from each pixel in Fig. \ref{fig:figure1}(h) and set all negative pixels to 0. The resulting PAD is shown in Fig. \ref{fig:figure2}(a).

In fact, as it will be seen from the results of the DFT simulation, some additional features due to photoemission from molecules are expected in the interstitial area between the blobs. These features are characterized by intensity values on the order of $10\%$ of the maximal value of the blobs intensity. However, in experimental PAD in Fig. \ref{fig:figure1}(h), the ratio between the corresponding mean pixel value in the interstitial area and the maximal value of the blobs intensity is on the order of $30\%$. This intensity thresholding will obviously cut weak molecular features that should be present according to the DFT simulation (see Fig. 3(h)), but they lie below the noise level of the data.

\subsection{\label{sec:Phase retrieval algorithms}Phase retrieval algorithms}

The PAD shown in Fig. \ref{fig:figure2}(a) was then used as the sole input for the iterative phase retrieval procedure, the details of which were published in our previous work~\cite{kliuiev:2016}. In brief, the procedure consists of alternating cycles of the phase-constrained~\cite{harder:2010} hybrid input-output~\cite{fienup:1982} (PC-HIO) and the error reduction~\cite{fienup:1978, fienup:1982} (ER) algorithms. The object support was obtained using the shrinkwrap algorithm~\cite{marchesini:2003}. For that, the initial estimate of the object support was obtained by computing the inverse Fourier transform of the processed experimental PAD data, $I_\mathrm{s}$, convolving it with a Gaussian function (standard deviation $\sigma=3$ pixels), thresholding at $9\%$ of its maximum and setting the pixel values below the threshold to zero. In the last iteration of each ER cycle, the output object distribution was used to update the object support by convolving it with a Gaussian function and setting a threshold at $19\%$ of its maximum. The width of the Gaussian was initially set to $\sigma=2.5$ pixels and was reduced by $1\%$ at every support update. In total, we performed 1000 independent reconstruction rounds with different initial random phase distributions. Each reconstruction round consisted of 5 alternating cycles of 10 iterations of the PC-HIO algorithm, 5 iterations of the ER algorithm and an update of the support. At the end of ten cycles, each reconstruction was stabilized by 100 iterations of the ER algorithm~\cite{latychevskaia:2015, kliuiev:2016}. We selected only 10 $\%$ of the object distributions having the lowest error metric in the reciprocal space~\cite{fienup:1978, fienup:1982, fienup:1986} and averaged them~\cite{shapiro:2005, thibault:2006, latychevskaia:2013, latychevskaia:2015, kliuiev:2016}.

Because an orbital distribution $\psi'(x,y)$ and its duplicate shifted by $(x_0, y_0)$ pixels, $\psi'(x-x_0,y-y_0)$, have the same amplitude of the Fourier transform, the location of the object support, obtained with the shrinkwrap algorithm, was arbitrary and partially reconstructed orbital distributions were often not aligned with the support constraint~\cite{fienup:1986}. As a consequence, we faced difficulties with determining the geometrical centre of the averaged orbital distribution.

To solve this problem, we undertook an additional reconstruction series, in which we employed the following support constraint. (i) The amplitude distribution obtained after averaging was centered in the computational domain using the central symmetry considerations. (ii) The centered amplitude distribution was convoluted with a Gaussian function (standard deviation $\sigma=1.3$ pixels) and the resulting image was thresholded at $21\%$ of its maximum. The threshold value was determined empirically so that the object distribution was not inadvertently truncated during the course of the reconstruction and the algorithm still converged. (iii) The thresholded amplitude distribution was then symmetrized with respect to its geometrical centre and the pixel values below 1 were set to 0, thus giving us a new object support. The support was kept steady in the centre of the computational domain during the reconstruction. Similarly, we performed 1000 independent reconstruction rounds and each round consisted of $20$ iterations of the PC-HIO algorithm followed by $20$ iterations of the ER algorithm. Only $10\%$ of the reconstructions with the lowest error metric were selected and averaged.

\section{\label{sec:Results and discussion}Results and discussion}

\subsection{\label{sec:Results of the iterative reconstruction}Results of the iterative reconstruction}

The results of the reconstruction are shown in Fig. \ref{fig:figure2}. The spatial resolution in the object domain was estimated to be $\Delta r_{||}= \frac{2\pi}{N\Delta k_{||}}\approx 1.57$~\AA, where the size of the pixel in the reciprocal space, $\Delta k_{||}=0.0074$~\AA$^{-1}$, and the linear number of pixels, $N=540$ pixels, were set by the experimental conditions. The reconstructed amplitude and phase distributions of the pentacene HOMO, obtained using the shrinkwrap algorithm with the uncentered support, are shown in Figs \ref{fig:figure2}(b) and \ref{fig:figure2}(c), respectively. The amplitude distribution shown in Fig. \ref{fig:figure2}(b) was used to obtain the new object constraint as described in the methods section. The reconstructed amplitude and phase distributions of the pentacene HOMO, obtained using the PC-HIO and ER algorithms with the centered tight support, are shown in Figs \ref{fig:figure2}(d) and \ref{fig:figure2}(e), respectively. We note that the amplitude distribution in Fig. \ref{fig:figure2}(b) shows two nodal planes in the center of the support (marked by crossing solid lines), while in Fig. \ref{fig:figure2}(d) it shows only one nodal plane in the center of the support (marked by crossing dotted lines). This is an example of a reconstruction artifact when the the correct reconstruction can be achieved only using the centered tight support. As it will be seen from the comparison with the results of density functional theory calculations, the use of the centered tight support indeed eliminated the problem of a translated object distribution, allowed for correct determination of its geometrical centre, and delivered artifact-free amplitude and phase distributions.

\subsection{\label{sec:Density functional theory simulation}Density functional theory simulation}

\begin{figure*}
\includegraphics{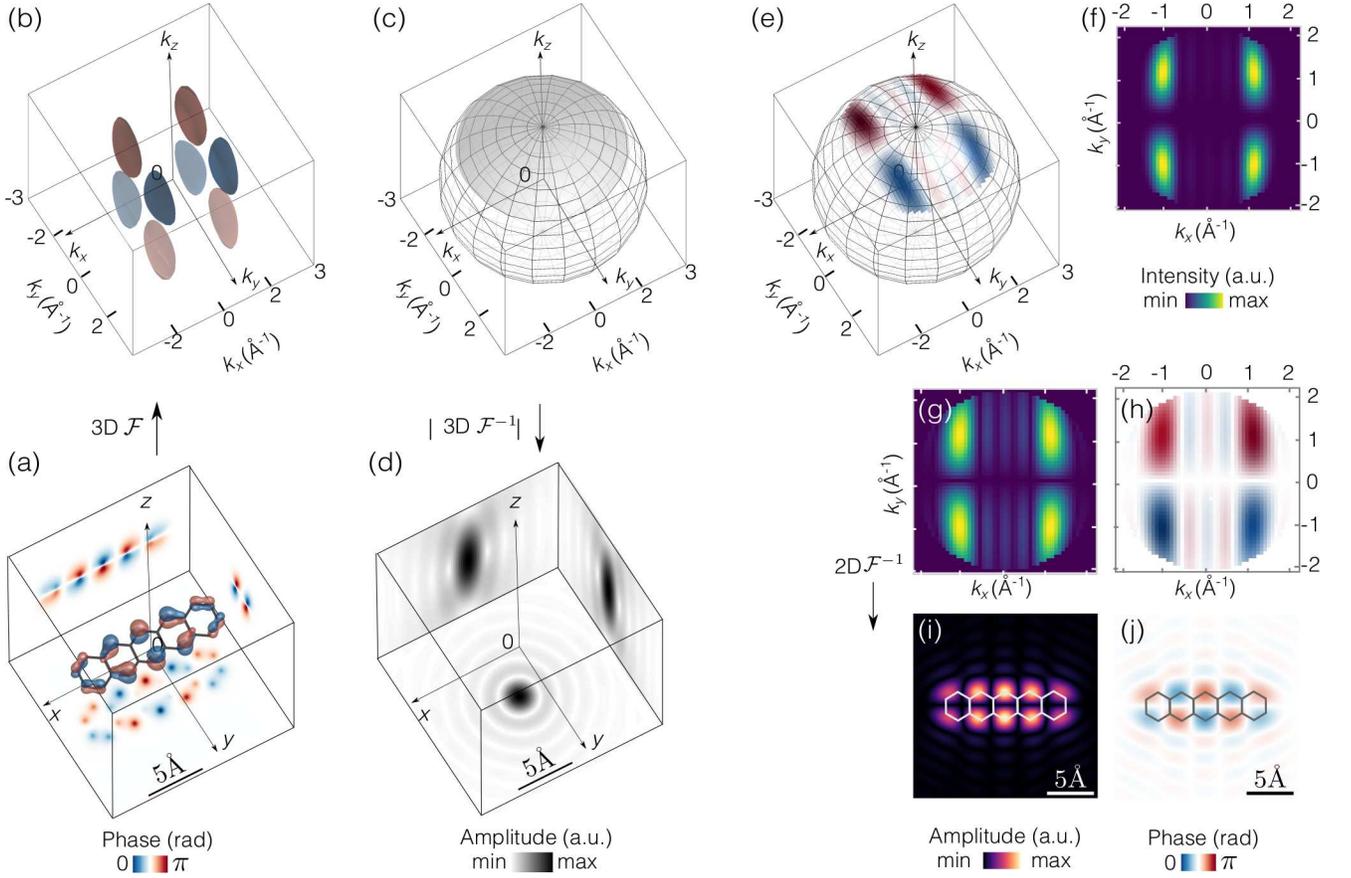}
\caption{(a-b) Results of the DFT simulation of the pentacene HOMO. (a) $\psi(x,y,z)$: 3D orbital distribution, represented as isosurface at the value of $50\%$ of maximum of $|\psi(x,y,z)|$. Inset images on the sides of the cube: Cross-sections through $\psi(x,y,z)$, computed at the planes located at $x\approx2.61$ \AA, $y\approx1.54$ \AA ~and $z\approx-0.15$ \AA. The phase values of the cross-sections are weighted with the corresponding amplitude values of $\psi(x,y,z)$ for illustration purposes.  (b) $\Psi(k_x,k_y,k_z)$: Fourier transform of $\psi(x,y,z)$, represented as an isosurface at the value of $50\%$ of the maximum of $|\Psi(k_x,k_y,k_z)|$. (c-j) Simulation of the experimental conditions using the results of the DFT simulation shown in (a-b). (c) Transfer function $H(k_x,k_y,k_z)$: $H=1$ on a segment of a hemisphere of radius $k_0=3.0$ \AA$^{-1}$ within the field of view of the parallel components of the momenta $k_{\mathrm{||, max}}=\pm 2$ \AA$^{-1}$, $H=0$ elsewhere.  (d) Inset images on the sides of the cube: Cross-sections through the amplitude of the response function $h(x,y,z)=\mathcal{F}^{-1}\{ H(k_x,k_y,k_z)\}$, computed at the planes located at $x=0$ \AA, $y=0$ \AA~ and $z=0$ \AA. (e) $\Psi(k_x,k_y,k_z)$ multiplied with the corresponding values of the transfer function $H$. (f) Squared modulus of the parallel projection of (e) onto the $(k_{x}, k_{y})$ plane. To be compared with the experimental PAD shown in Fig. \ref{fig:figure2}(a). (g) and (h) Amplitude and phase distributions of the parallel projection of (e) onto the $(k_{x}, k_{y})$ plane. (i) and (j) Amplitude and phase distributions in real space obtained by computing the inverse Fourier transform of (g) and (h). The phase values are weighted with the corresponding amplitude values for illustration purposes. To be compared with the reconstructed orbital distributions shown in Fig. \ref{fig:figure2}(d) and (e). In (a,d), the gray borders of the cubes mark $110\times110\times 110$ pixels sections cut out from $512\times 512 \times 512$ pixels DFT data. In (b,c,e), the gray borders of the cubes mark $75\times75\times 75$ pixels sections cut out from $512\times 512 \times 512$ pixels DFT data.}
\label{fig:figure3}
\end{figure*}

Electronic structure calculations of the pentacene HOMO in free space were performed using Kohn-Sham density functional theory (DFT)~\cite{hohenberg:kohn:1964, kohn:1965} at the PBEPBE/cc-pVDZ level as implemented in the $Gaussian$ quantum chemistry suite~\cite{gaussian}. The three-dimensional (3D) orbital distribution was centered in the computational domain. The size of the pixel was $\Delta r=0.1538~\mathrm{\AA}$ in each dimension and the domain was sampled at $N_{x_0}\times N_{y_0} \times N_{z_0}=51\times140\times72$ pixels. The 3D orbital distribution was zero-padded to $N=512$ pixels in each dimension, yielding the orbital distribution $\psi(x,y,z)$. Fig. \ref{fig:figure3}(a) shows an orbital isosurface plotted at $50\%$ of the maximal value of $|\psi(x,y,z)|$ and three slices made at selected planes. The Fourier transform of $\psi(x,y,z)$ delivered the distribution $\Psi(k_x,k_y,k_z)$. Figure \ref{fig:figure3}(b) shows an isosurface plotted at $50\%$ of the maximal value of $|\Psi(k_x,k_y,k_z)|$. The effective size of the pixel in the reciprocal space was equal to $\Delta k=\frac{2\pi}{N\Delta r}\approx 0.08$ \AA$^{-1}$. 

To provide means for the quantitative comparison of the DFT results with the results of the iterative reconstruction and the experiment, we did a simulation by modifying the distribution $\Psi(k_x,k_y,k_z)$ so that it complied with the experimental conditions. (i) In the experiment, the PAD was recorded at fixed kinetic energy. As optical transitions are direct transitions in reciprocal space, the momenta in the PAD are sampled on a hemisphere~\cite{puschnig:2009, puschnig:2013, weiss:2015} with radius $k_0$ set by the photoelectron kinetic energy to $k_0= 0.512\cdot\sqrt{E_{\mathrm{kin}}}$. At $E_{\textrm{kin}}=34.3$ eV, the radius was $k_0\approx3.0$ \AA$^{-1}$. Thus, all pixel values in the range of $[0.95k_0,~1.05k_0]$ were kept and all others were set to 0. (ii) The values of all pixels lying outside of the $|k_{\mathrm{||}}|_\mathrm{max}=2$ \AA$^{-1}$ range were set to 0, in order to account for the numerical aperture of the electron entrance optics of the PEEM. The choice of the $[0.95k_0,~1.05k_0]$ range was justified by two reasons. First, due to a limited number of pixels per reciprocal length unit, keeping the pixel values in this range ensured that no pixels were missing on the surface of the hemisphere. Second, this accounted for the (lorentzian) broadening of the photoelectron momentum perpendicular to the surface due to the exponential decay of the electron wave function inside the solid~\cite{baumberger:2004}. The decay length is essentially given by the inelastic mean-free path, which is of the order of 1 nm. This decay translates in a lorentzian momentum broadening of about $0.1$ \AA$^{-1}$, which agrees well with the interval chosen above. 

The modification steps described above corresponds to multiplication of the $\Psi(k_x,k_y,k_z)$ distribution with a transfer function $H=H(k_x,k_y,k_z)$:
\begin{equation}\label{eq:transfer_function}
    H= 
\begin{cases}
   1 & \text{for } k_z=\sqrt{k_0^2-k_x^2-k_y^2} \text{ }  \cap \text{ }  k_x^2+k_y^2 \leq k_{\mathrm{||, max}}^2\\
    0              & \text{elsewhere}.
\end{cases}
\end{equation}

It represents a thin segment on the $k$-sphere as plotted in Fig. \ref{fig:figure3}(c). Cross-sections through the amplitude of its complex-valued inverse Fourier transform done at selected planes are shown in Fig. \ref{fig:figure3}(d). The resulting complex-valued distribution $\Psi_H\left(k_x,k_y,\sqrt{k_0^2-k_x^2-k_y^2}\right)=\Psi(k_x,k_y,k_z) H(k_x,k_y,k_z)$ is shown in Fig. \ref{fig:figure3}(e). The squared modulus of this distribution, projected parallelly onto the $(k_x,k_y)$ plane, shown in Fig.~\ref{fig:figure3}(f), corresponds to the intensity distribution measured in the experiment and is in good agreement with the PAD shown in Fig.~\ref{fig:figure2}(a). Amplitude and phase of $\Psi_H\left(k_x,k_y,\sqrt{k_0^2-k_x^2-k_y^2}\right)$ are shown in Fig. \ref{fig:figure3}(g,h). The inverse Fourier transform of the complex-valued distribution $\Psi_H\left(k_x,k_y,\sqrt{k_0^2-k_x^2-k_y^2}\right)$ yielded a 2D orbital distribution in real space:

\begin{eqnarray}
\psi'(x,y)=\iint \limits_{-|k_{\mathrm{||}}|_\mathrm{max}}^{+|k_{\mathrm{||}}|_\mathrm{max}}
\Psi_H\left(k_x,k_y,\sqrt{k_0^2-k_x^2-k_y^2}\right) \cdot \nonumber \\  \cdot e^{ik_xx+ik_yy}dk_xdk_y,
\label{eq:2D_IFT}
\end{eqnarray}
where the prime symbol distinguishes it from the original DFT data. Amplitude and phase of $\psi'(x,y)$ are shown in Fig. \ref{fig:figure3}(i,j). These distributions mathematically correspond to those obtained by the iterative reconstruction  shown in Fig. \ref{fig:figure2}(d,e). We find them to be in a good agreement, with some minor differences in the shapes of the individual lobes. The larger spatial extent of the reconstructed orbital distribution in Fig. \ref{fig:figure2} can be attributed to the side effects of the image processing procedure: in the processed experimental PAD, shown in Fig. \ref{fig:figure2}(a), the blobs are more confined than those in the 2D distribution shown in Fig. \ref{fig:figure3}(f) and the features in the interstitial area between the blobs disappear after the processing because they lie below the noise level. Consequently, the reconstructed orbital distribution becomes more delocalized in space. 

\subsection{\label{sec:Interpretation of 2D orbital distributions and 3D reconstruction from 2D experimental data}Interpretation of 2D orbital distributions and 3D reconstruction from 2D experimental data}

In Eq. \ref{eq:2D_IFT}, we employed the inverse Fourier transform to compute the 2D orbital distribution $\psi'(x,y)$ from the complex-valued photoelectron distribution $\Psi_H\left(k_x,k_y,\sqrt{k_0^2-k_x^2-k_y^2}\right)$. In essence, the same equation was employed while reconstructing the orbital distributions from the photoemission data. However, following the considerations of Barton~\cite{barton:1988}, a more general expression for the orbital distribution $\psi'$ is appropriate:
\begin{eqnarray}
\psi'(x,y,z)=\iint \limits_{-|k_{\mathrm{||}}|_\mathrm{max}}^{+|k_{\mathrm{||}}|_\mathrm{max}} \Psi_H\left(k_x,k_y,\sqrt{k_0^2-k_x^2-k_y^2}\right) \cdot  \nonumber \\ \cdot e^{iz\sqrt{k_0^2-k_x^2-k_y^2}}e^{ik_xx+ik_yy}dk_xdk_y,
\label{eq:bartons_integral}
\end{eqnarray}
which we will refer to as ``Barton's integral''~\cite{barton:1988}. The resulting reconstruction at $z=0$ is shown in Fig.  \ref{fig:figure3}(i,j). If the integral in Eq. \ref{eq:bartons_integral} is computed at some $z=z_0=\mathrm{const}$, then the resulting 2D orbital distribution $\psi'(x,y)$ is a superposition of the ``in-focus'' contribution $\psi(x,y,z_0)$ and  the ``out-of-focus'' signal from adjacent z-planes as defined by the axial resolution of the experimental system.

The range of $z$ planes contributing to the ``out-of-focus'' component can be estimated using the formula for the axial resolution defined by the Rayleigh range~\cite{gross2:2005}:
\begin{equation}\label{eq:axial_resolution}
\delta z=\frac{2\lambda}{\mathrm{NA}^2}.
\end{equation}

Given the de Broglie wavelength of electrons at $34.4$ eV kinetic energy is $\lambda\approx2.1$ \AA~ and the numerical aperture defined by the geometry in Fig. \ref{fig:figure3}(c) is $\mathrm{NA}=\frac{|k_{\mathrm{||, max}}|}{k_0}=0.67$, the axial resolution is $\delta z\approx9.4$ \AA. The transverse resolution can be computed using the formula for the Airy radius~\cite{gross2:2005}:
\begin{equation}\label{eq:transverse_resolution}
\delta r_\mathrm{||}=\frac{0.61\lambda}{\mathrm{NA}},
\end{equation}
giving $\delta r_\mathrm{||}\approx 1.9$ \AA. These results are in a good agreement with the corresponding values estimated by computing the square of the amplitude of the response function $h(x,y,z)$ shown in Fig. \ref{fig:figure3}(d). 

Another important point we would like to highlight is the following. From the phase retrieval, we recover the complex-valued distribution $\Psi_H\left(k_x,k_y,\sqrt{k_0^2-k_x^2-k_y^2}\right)$. Now we note that in Eq. \ref{eq:bartons_integral}, the factor $e^{iz\sqrt{k_0^2-k_x^2-k_y^2}}$ plays the role of a propagator along the z-dimension of the orbital distribution. Thus, by computing Barton's integral at various values of $z$, one gains access to different ``in-focus'' contributions of the 3D orbital distribution $\psi(x,y,z)$ estimated at the planes $z$ and one could thereby reconstruct the full 3D orbital distribution solely from the 2D distribution $\Psi_H\left(k_x,k_y,\sqrt{k_0^2-k_x^2-k_y^2}\right)$! This remarkable result was obtained by Barton in a simulated example of the holographic reconstruction of a $\mathrm{S(1s)}$ photoemitter signal from $c(2\times2)\mathrm{S/Ni(001)}$ data~\cite{barton:1988}. Though, for flat molecules with the thickness on the order of $2...3$ \AA, this will require improvement of axial resolution by increasing photon energies and/or numerical aperture of the system.

Using these results, we can also elaborate on reasons why orbital tomography was so far applied to planar molecules only~\cite{puschnig:2009, puschnig:2013}. From one point of view, this limitation can be justified by the need of absence of scatterers on the way of the photoelectron wave as it propagates upon excitation from the molecule to the detector. However, another point of view can be gained from the interpretation of the 2D reconstruction as a superposition of ``in-focus'' and ``out-of-focus'' contributions. If a molecular orbital distribution is planar, the 2D reconstruction will contain ``in-focus'' and ``out-of-focus'' contributions having similar patterns in different $z$ planes, blurred in accordance with the corresponding depth of field. The 2D orbital distribution thus represents a good estimate of the 3D orbital distribution. Otherwise, if the molecules are, for example, non-planar, the 2D reconstruction will contain ``in-focus'' and ``out-of-focus'' contributions of very different patterns from different $z$ planes. The corresponding 2D reconstruction alone is thus no longer a good estimate of the 3D orbital distribution, but only some effective distribution defined by the orbital geometry and the depth of field. Unambiguous 3D reconstruction will thus require measuring PADs at multiple photon energies and solving the 3D phase problem, provided the oversampling requirements~\cite{miao:1998, miao:2003} are fulfilled in all dimensions.

\section{\label{sec:Summary and conclusion}Summary and conclusion}

To facilitate the pre-processing of the experimental data for their use in the phase retrieval algorithm, we proposed an image processing procedure, which is based on the maximization of mutual information algorithm and allows for efficient subtraction of the background features. We employed this procedure to process the experimental ARPES data recorded from pentacene on Ag(110) and we were able to successfully reconstruct both the amplitude and phase distribution of the highest occupied molecular orbital by means of our phase retrieval routine. The quality of the reconstruction was improved by introducing the second reconstruction run, in which the centered and thresholded reconstructed object distribution obtained after the first reconstruction run was used as a tight object support. This eliminated the ambiguity about the location of the geometrical centre of the orbital distribution and improved the overall contrast of the reconstructed data. The results of the reconstruction were then compared with the DFT simulation, obtained by modifying the original 3D DFT data in accordance with the experimental conditions. Both reconstructed and DFT data revealed good agreement. 

The reconstructed 2D orbital distributions obtained in phase retrieval can be viewed as a superposition of the ``in-focus'' orbital distribution at the plane $z=0$ and ``out-of-focus'' orbital distributions at other planes $z=\mathrm{const}$ whose strength and blur are set by the depth of focus of the experimental arrangement.  Most importantly, we came to conclusion that by computing the integral of Barton, one can reconstruct full 3D orbital distribution solely from a single set of the complex-valued 2D distributions in reciprocal space, the phase of which was obtained in the phase retrieval procedure. Unambiguous 3D reconstruction will require photoemission data acquired at several photon energies though.

\begin{acknowledgments}
Financial support by the Swiss National Science Foundation through NCCR MUST is gratefully acknowledged. The authors thank Vitaliy Feyer for helpful discussions and his support during the beamtime at the NanoESCA beamline at the Elettra synchrotron. 
\end{acknowledgments}

\appendix*
\section{\label{sec:appendix} }

\subsection{Experimental geometry and calibration}

\begin{figure}
\includegraphics{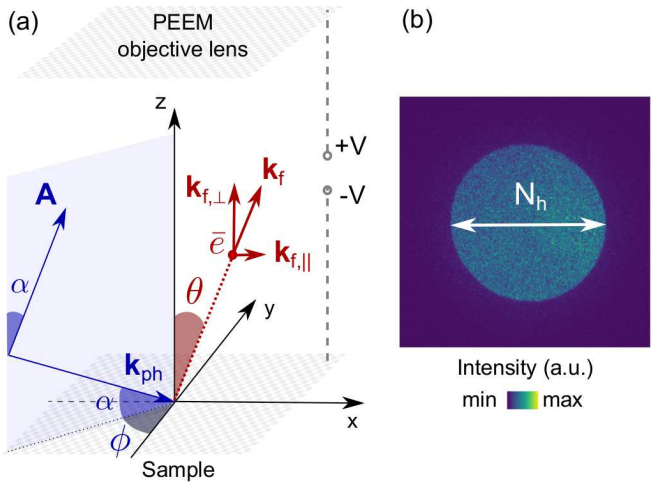}
\caption{\label{fig:figure4} (a) Experimental geometry. The 40 eV p-polarized light was incident on the sample at a grazing angle of $\alpha=25^\circ$. The azimuthal angle between the plane of incidence and the $[1\overline{1}0]$ high symmetry direction of the Ag(110) crystal was $\phi=5^\circ$. The vector potential and wave vector of light are denoted by $\textit{\textbf{A}}$ and $\textit{\textbf{k}}_\mathrm{ph}$, respectively. The photoelectrons were collected by the PEEM objective lens. $\textit{\textbf{k}}_{\mathrm{f},\parallel}$ and $\textit{\textbf{k}}_{\mathrm{f},\perp}$ denote parallel and normal components of the final state wave vector of the photoelectrons. (b) Secondary photoelectron horizon recorded at 7 eV photoelectron kinetic energy. The full width of the horizon, $N_h$, corresponds to $2|\textit{\textbf{k}}_\mathrm{\mathrm{f}, ||} ^{\mathrm{max}}|\approx 2.7$ \AA$^{-1}$ and was equal to $N_\mathrm{h}=365\pm4$ pixels.}
\end{figure}

The experimental geometry was identical to that described in our earlier work~\cite{kliuiev:2016} and is shown in Fig. \ref{fig:figure4}(a).
The scale in reciprocal space was calibrated using a photoelectron horizon of the secondary electrons emitted from the clean Ag(110) sample upon excitation with the p-polarized light of 40 eV photon energy. At $E_{\mathrm{kin}}=7$ eV, the maximal value of the parallel component of the final state wave vector was $|\textit{\textbf{k}}_\mathrm{\mathrm{f}, ||} ^{\mathrm{max}}|\approx 1.35$ \AA$^{-1}$.  The full width of the horizon $2|\textit{\textbf{k}}_\mathrm{\mathrm{f}, ||} ^{\mathrm{max}}|$ was $N_\mathrm{h}=365\pm4$ pixels as it is shown in Fig. \ref{fig:figure4}(b). The pixel size in reciprocal space was then determined to be $\Delta k_\mathrm{||}=0.0074~\mathrm{\AA}^{-1}$.

\subsection{Registration procedure}

In order to obtain sufficiently high signal-to-noise ratio of the pentacene valence state PAD, we acquired data in a 200 meV energy window, which is of the order of the electron analyser resolution and of the full-width at half-maximum of the pentacene HOMO at the binding energy of 1.2 eV. The data were taken in steps of 20 meV with the dwell time of 3 s per image. In total, we did 5 independent recordings of the data in this energy window and summed up the resulting 50 PADs. Because organic molecules suffer from radiation damage under intense UV and x-ray irradiation, the PADs were acquired with short acquisition times and raster scanning of the sample. This caused image drift and, as a result, the normal emission direction in the individual PADs was not aligned to a common pixel on the CCD. As a result, we had to perform numerical registration of the data, i.e. the overlaying of all 50 PAD images acquired at different times with the purpose of their geometrical alignment~\cite{zitova:2003}.  The mutual translation, $\bf{\Delta}$, of the experimental PADs, $I=I(\textit{\textbf{k}}_{\mathrm{f}\parallel}, E_{\mathrm{kin}})$, sampled at $N\times N=540\times 540$ pixels, was only on the order of 8 pixels both in horizontal and vertical directions. Therefore, for registration of the PAD data, we choose the method of intensity interpolation based on the cross-correlation~\cite{tian_huhns:1986}. This method was found to outperform all other relevant registration methods~\cite{brown:1992} in case of a real-valued image sequence contaminated with noise~\cite{tian_huhns:1986}. To find the amount of shift, the experimental PADs were (i) up-sampled by a factor of $s=2$ to $N'\times N'=1080\times1080$ pixels by bicubic interpolation and (ii) normalized as $ I'_\mathrm{norm}(u,v) = I'(u,v)/ \sqrt{\sum_{u,v=0}^{N'-1}I'(u,v)}$, where $u$ and $v$ are the coordinates in the detector plane and the prime symbol denotes the up-sampling. In addition, (iii) the mean value was subtracted from each pixel of the normalized images~\cite{dvornychenko:1983}. (iv) The position of the cross-correlation maximum of the first two up-sampled images, $\bf{\Delta}'$, delivered the relative shift between the experimental PAD images, $\bf{\Delta}=\bf{\Delta}'$$/s$, which was used to merge them by summation. (v) The merged image was then used as a reference to bring it into registry with the next image. (vi) The process was repeated sequentially until the whole image sequence was aligned. The PAD obtained upon registration of 50 data sets is shown in Fig. \ref{fig:figure1}(a). 

\subsection{Normalization with $|\textit{\textbf{A}}\cdot \textit{\textbf{k}} _{\mathrm{f}} |^2$ factor}

After the removal of the background features, the PAD intensity distribution had to be normalized. As it is seen from Fig. \ref{fig:figure4}(a), the angle of incidence of the incoming light, $\alpha=25^\circ$, was kept fixed, while the photoelectrons were detected at a broad range of polar angles $\theta$. Thus, the intensity distribution in the detector plane became modulated by the $|\textit{\textbf{A}}\cdot \textit{\textbf{k}} _{\mathrm{f}} |^2$ factor, as set by Eq.~\ref{eq:photocurrent}. This modulation was leveled by dividing the PADs shown in~Fig. \ref{fig:figure1}(a,c,d) by
\begin{eqnarray*}
&&|\textit{\textbf{A}}\cdot \textit{\textbf{k}} _{\mathrm{f}} |^2=|A_xk_{\mathrm{f}, x}+A_yk_{\mathrm{f}, y}+A_zk_{\mathrm{f}, z}|^2= \nonumber\\
&&=\left|A_xk_{\mathrm{f}, x}+A_yk_{\mathrm{f}, y}+A_z\sqrt{\frac{2m}{\hbar^2}(E_\mathrm{kin}+V_0)-k_{\mathrm{f}, x}^2-k_{\mathrm{f}, y}^2}\right|^2,
\end{eqnarray*}
where the electric field vector potential components $A_x=\sin\alpha \cdot \sin \varphi$, $A_y=\sin\alpha \cdot \cos\varphi$, $A_z=\cos\alpha$ and the polar and azimuthal angles of incidence were defined by the experimental geometry. The mean inner potential, $V_0$, which typically varies between 5 to 9 eV for overlayers of common organic molecules~\cite{lunt:2009, greif:2013}, was set to $V_0=7$ eV, as this value led to the best results in terms of the symmetry of the corrected data.  The resulting normalized PADs are shown in Fig. \ref{fig:figure1}(f-h).

\bibliography{biblio.bib}

\end{document}